\documentclass[12pt]{JHEP3}
\usepackage{epsfig}
\usepackage{amsmath}
\newcommand{\unit}{\hbox to 3.8pt{\hskip1.3pt \vrule height 7.4pt
    width .4pt \hskip.7pt \vrule height 7.85pt width .4pt \kern-2.4pt 
    \hrulefill \kern-3pt \raise 3.7pt\hbox{\char'40}}}
\newcommand{\p}{\partial}
\newcommand{\Tr}{{\rm Tr}}
\newcommand{\nn}{\nonumber}
\newcommand{\ket}[1]{{\left|#1\right\rangle}}
\newcommand{\bra}[1]{{\left\langle#1\right|}}
\title{
Stringy Derivation of Nahm Construction of Monopoles
}
\author{
Koji Hashimoto$^\dagger$ and Seiji Terashima$^*$\\ 
$^\dagger$Institute of Physics, the University of Tokyo,\\
\hspace{5mm}3-8-1 Komaba, Tokyo 153-8902, Japan\\
\hspace{2mm}The Fields Institute for Research in Mathematical Sciences,\\
\hspace{5mm}222 College Street, Toronto, ON M5T 3J1, Canada\\
\hspace{2mm}E-mail: \email{koji@hep1.c.u-tokyo.ac.jp}\\
$^*$New High Energy Theory Center, Rutgers University,\\ 
\hspace{5mm}126 Frelinghuysen Road, Piscataway, NJ 08854-8019, U.S.A.\\
\hspace{2mm}E-mail: \email{seijit@physics.rutgers.edu}\\
}

\abstract{
We derive the Nahm construction of monopoles from 
exact tachyon condensation on unstable D-branes.
The Dirac operator used in the Nahm construction is identified
with the tachyon profile in our D-brane approach, and 
we provide physical interpretation of the procedures Nahm gave.
Crucial is the introduction of infinite number of brane-antibranes
from which arbitrary D-brane can be constrcuted, exhibitting a unified
view of various D-branes.
We explicitly show the equivalence of the D3-brane boundary state with
the monopole profile and the D1-brane boundary state with the Nahm data
as transverse scalars. 
}

\preprint{
{\normalsize{\tt hep-th/0507078}}\\ 
{\normalsize UT-Komaba/05-6}
}

\begin{document}

\section{Introduction and Summary of the Derivation}

D-branes, owing to their variety in dimensions and protean shapes, 
provide new accesses to nonperturbative aspects in field theories.
In particular, solitons in field theories often admit stringy
interpretation, which enables one to view them as geometric objects: 
compound branes in higher dimensions. Once the corresponding 
brane configuration is identified, one can make various
stringy deformations/dualities ending up with new perspectives on the
solitons. For example, different kinds of D-branes are related and
equivalent, especially through tachyon condensation \cite{senconj} 
describing creation/annihilation of D-branes. 
For a demonstration of the powerfulness of the D-brane technique, 
study of instantons/monopoles is appropriate due to their obvious
importance. In this article, we {\it derive} completely the Nahm's
construction of BPS monopoles 
\cite{Nahm:1979yw,Corrigan:1983sv}, using a D-brane technique
in string theory --- exact tachyon condensation. For the derivation
we apply the idea of K-matrix theory \cite{kmatrix,Asakawa:2002ui} 
in which any species of D-branes can be described by the tachyon
condensation of infinite number of a single kind of D-branes and
anti-D-branes. 

\begin{figure}[tp]
\begin{center}
\begin{minipage}{13cm}
\begin{center}
\includegraphics[width=6cm]{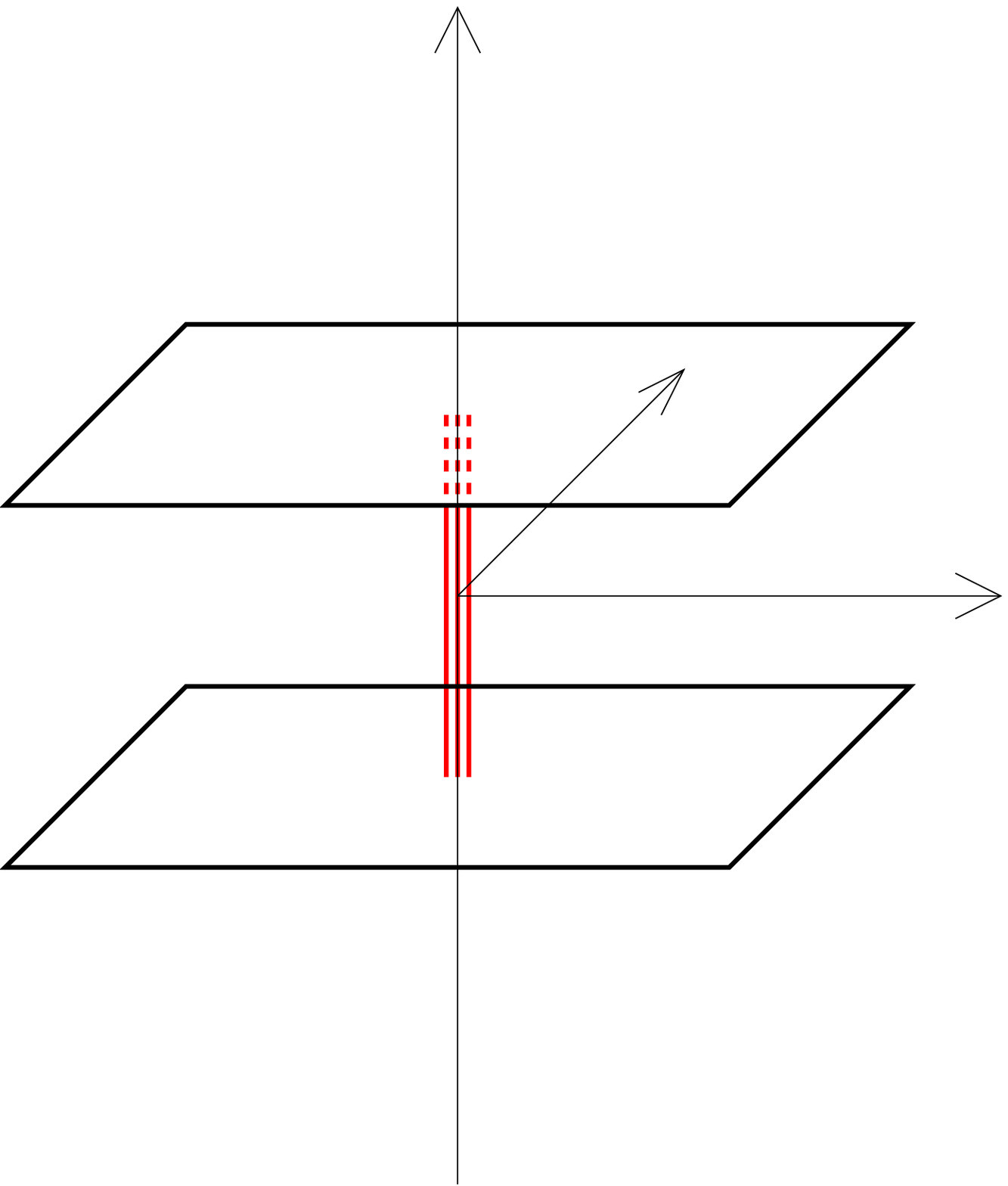}
\put(-130,70){D3}
\put(-130,130){D3}
\put(-115,100){D1}
\put(-105,190){$\xi$}
\put(-5,110){$x^1,x^2,x^3$}
\caption{A schematic picture of the brane configuration corresponding to
 typical BPS monopoles in SU(2) super Yang-Mills theory.}
\label{fig0}
\end{center}
\end{minipage}
\end{center}
\end{figure}

Brane configuration of $k$ BPS monopoles in SU($N$) super Yang-Mills 
theory is composed of $k$ D-strings (D1-branes)
suspended between parallel $N$
D3-branes (Fig.\ref{fig0}) in type IIB superstring theory. 
Evidence for this correspondence is through
identification of masses, charges and supersymmetries preserved
\cite{Green:1996qg}. The equations for the monopole
\begin{eqnarray}
 D_i \Phi = \frac{i}{2} \epsilon_{ijk} F_{jk} \, \quad 
(i,j,k=1,2,3)
\label{bpseqmon}
\end{eqnarray}
can be viewed as BPS equations in the effective field theory 
on the D3-branes. The brane interpretation made it possible to predict
existence of generalized monopoles, such as monopoles in noncommutative
spaces \cite{ncmono} and 1/4 BPS dyons \cite{1/4bpsdyon}, 
whose explicit field configurations were obtained later accordingly
\cite{solutions,solutions2}. In the Nahm's method,
arbitrary BPS monopoles can be constructed through definite procedures
starting with solving Nahm's equations 
\begin{eqnarray}
 \p_\xi T_i(\xi) = i \epsilon_{ijk} T_j(\xi) T_k(\xi)
\label{nahmeq}
\end{eqnarray}
to get their $k\times k$ matrix solutions $T_i(\xi)$ called Nahm data,
as briefly reviewed below. It was shown by Diaconescu 
\cite{Diaconescu:1996rk} that 
the Nahm's equations are in fact BPS equations in the worldvolume
effective 
theory on the D-strings. These already exhibit how powerful the D-brane
technique is, but it was not enough, since D-brane interpretation of the
Nahm's construction itself has not been revealed, except for 
the partial derivation by the probe analysis \cite{Diaconescu:1996rk}.
We shall show that in fact {\it all} the procedures Nahm provided for
the construction have stringy physical interpretation in higher
dimensions, and resultantly derive the Nahm's monopole construction
which was obtained originally from ADHM construction
\cite{Atiyah:1978ri}.  

\begin{figure}[tp]
\begin{center}
\begin{minipage}{5cm}
\includegraphics[width=5cm]{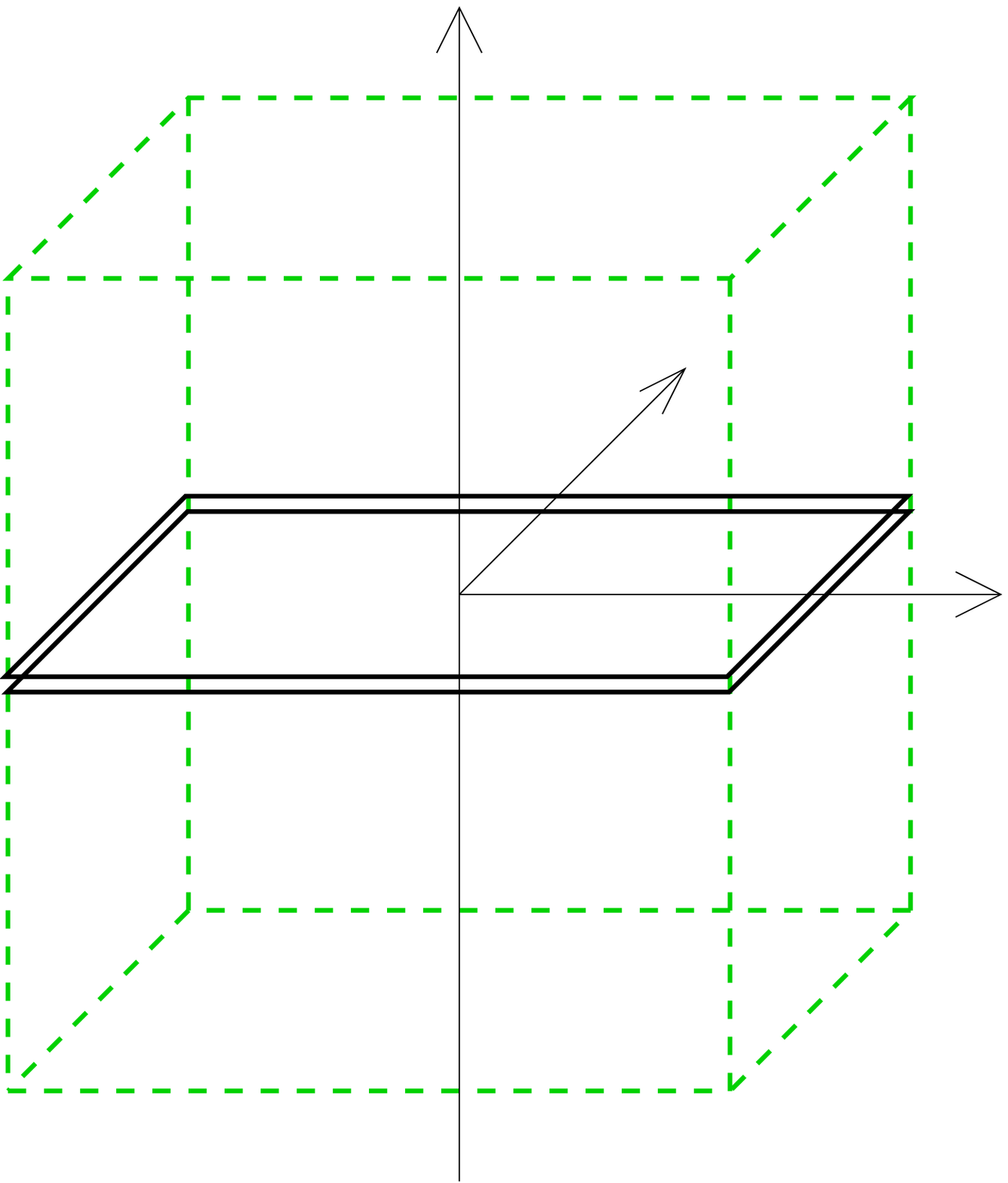}
\put(-109,58){D3}
\put(-50,45){Non-BPS D4}
\put(-89,159){$\xi$}
\put(5,82){$x^i$}
\end{minipage}
\hspace{1cm}
\begin{minipage}{5cm}
\includegraphics[width=5cm]{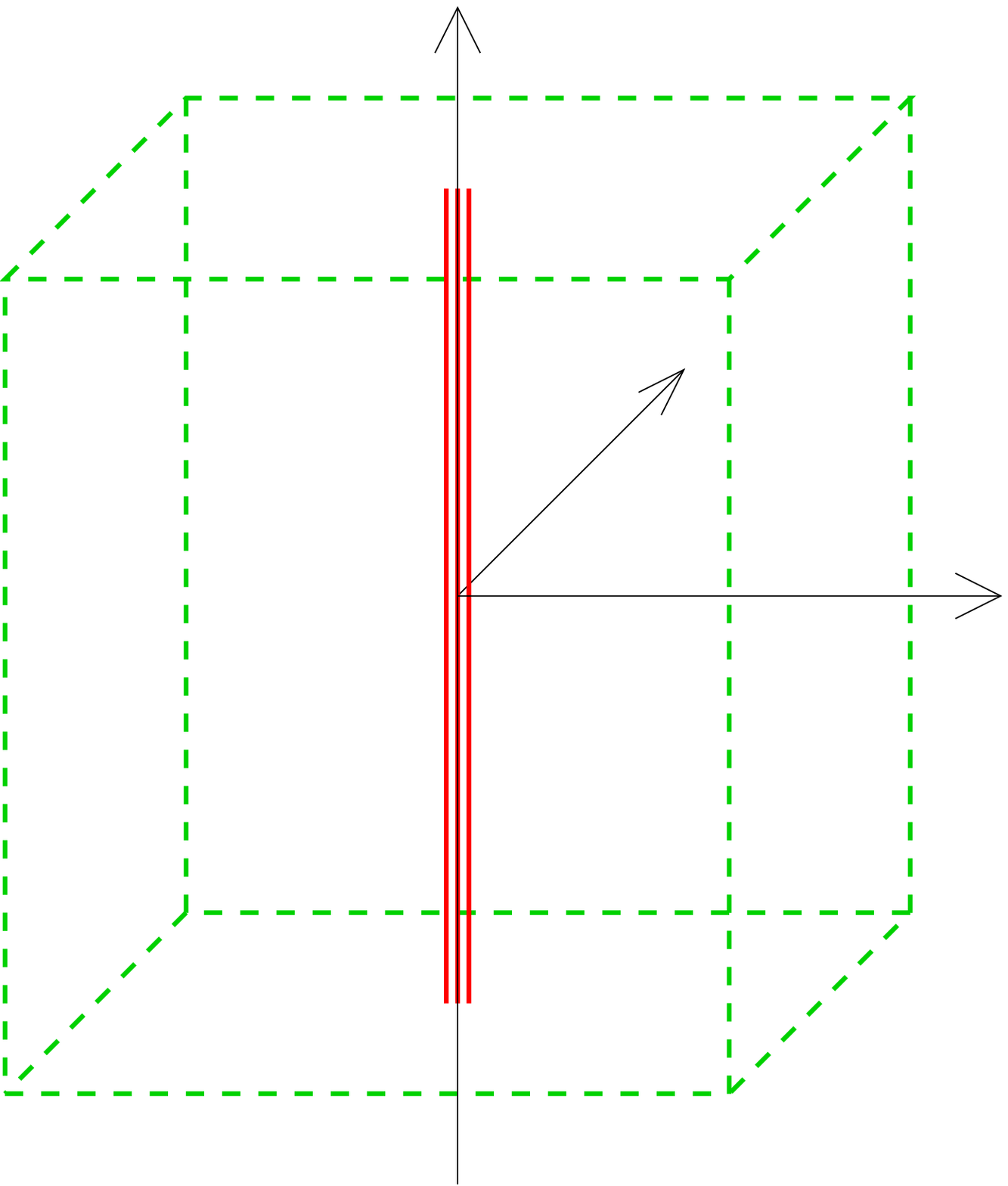}
\put(-97,84){D1}
\put(-50,45){Non-BPS D4}
\put(-89,159){$\xi$}
\put(5,82){$x^i$}
\end{minipage}
\begin{center}
\begin{minipage}{13cm}
\caption{Left: D3-branes and anti-D3-branes are shown in solid planes,
 while the dashed box represents non-BPS D4-branes. Right: Solid lines
 are D-strings made out of the non-BPS D4-branes (the dashed box) by the 
 tachyon condensation.}
\end{minipage}
\end{center}
\label{fig2}
\end{center}
\end{figure}

Nahm argued \cite{Nahm:1979yw} that, with the solutions to
the equations (\ref{nahmeq}) in a period $\xi\in (-a,a)$ with a certain
boundary condition, normalizable wave functions 
$\bra{\xi}\rho,x^i\rangle$
of the ``Dirac operator'' (in one dimensional space spanned by $\xi$)
\begin{eqnarray}
 \nabla \equiv \unit_{2k\times 2k}\frac{\p}{\p \xi} 
+ i\sigma_i \otimes (x^i \unit_{k\times k} - T_i(\xi)), 
\qquad 
 \nabla\ket{\rho,x^i}=0 \quad (\rho=1,2,\cdots,N)\quad 
\end{eqnarray}
provide monopole field profiles by the formula
\begin{eqnarray}
&&\Phi(x^i)_{\rho\rho'}
\equiv \bra{\rho, x^i} \xi \ket{\rho', x^i},
\label{nahmformp}\\
&&A_i(x^j)_{\rho\rho'}
\equiv \bra{\rho, x^j} \frac{\p}{\p x^i} \ket{\rho',x^j}.
\label{nahmforma}
\end{eqnarray}
To derive these full procedures in string theory, we resort to the 
idea of tachyon condensation which provides a unified viewpoint of
the D3-branes and the D-strings. The K-matrix theory \cite{kmatrix}
ensures that infinite number of D$p$-branes and anti-D$p$-branes 
(or non-BPS D$p$-branes)
can reproduce any kind of D-branes, and since we are interested in
monopoles living on the D3-brane worldvolume theory, let us start with
infinite number of parallel D3-branes and anti-D3-branes along the 
$x^i$ directions. To see the relation to the D-string picture on which
the Nahm data is realized \cite{Diaconescu:1996rk}, we consider on the
D3-branes a tachyon condensation representing the D-strings. 
The D-string worldvolume direction $\xi$ is orthogonal to
the worldvolume direction of the D3-brane $(x^1, x^2, x^3)$, thus 
we are inevitably lead to introduce non-BPS D4-branes whose worldvolume
spans $(x^1,x^2,x^3,x^4\equiv \xi)$, as an intermediate step. 
See Fig.\ref{fig2}(Left) in which the dashed box is the non-BPS
D4-branes (on top of each other). 
Once we construct the non-BPS D4-branes from the D3-branes by the
tachyon condensation, then we further make another tachyon condensation 
to realize the D-strings, as in Fig.\ref{fig2}(Right). 
Therefore the tachyon condensation is a sequence 
\begin{eqnarray}
\left(
\begin{array}{c}
 \mbox{D3-branes}\\
\mbox{anti-D3-branes}
\end{array}
\right)
  \quad\rightarrow\quad \mbox{non-BPS D4-branes}
\quad\rightarrow\quad \mbox{D-strings}.
\label{seq}
\end{eqnarray}

\begin{figure}[jtp]
\begin{center}
\begin{minipage}{13cm}
\begin{center}
\includegraphics[width=6cm]{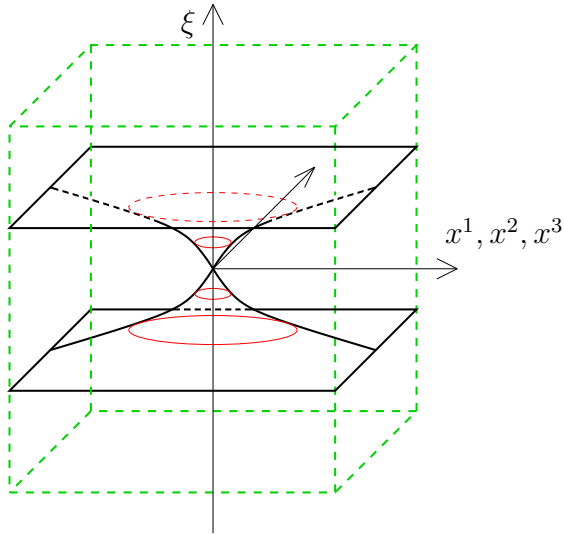}
\put(-105,190){$\xi$}
\put(-5,110){$x^1,x^2,x^3$}
\caption{The actual brane configuration representing the BPS monopoles
in the SU(2) gauge theory.}
\label{fig1}
\end{center}
\end{minipage}
\end{center}
\end{figure}

Although schematically the brane configuration corresponding to the 
monopoles is Fig.\ref{fig0}, the actual configuration looks like 
Fig.\ref{fig1}. The worldvolumes of the D3-branes are 
curved due to the excitation of the transverse scalar field $\Phi(x^i)$.
There is another way of interpreting the picture of Fig.\ref{fig1}: 
by turning on the transverse scalar fields $T_i(\xi)$
on the D-strings of Fig.\ref{fig2}(Right), 
the desired brane configuration Fig.\ref{fig1} can be
constructed, in which the D-string worldvolume is expanded into 4
dimensions via renowned Myers' effect \cite{Myers}. This 
Myers' effect occurs for mutually-noncommutative matrix transverse 
fields on the D-strings, which is manifested in the Nahm's equation
(\ref{nahmeq}). 
In summary, to produce the brane configuration Fig.\ref{fig1}, we have
two ways --- 
\begin{itemize}
 \item[$\langle$i$\rangle$] 
The D3-branes with the monopole field configuration which
should be a direct interpretation of the starting entry of the
sequence (\ref{seq}).
\item[$\langle$ii$\rangle$]  The expanded D-strings with the Nahm data,
the end point of the sequence (\ref{seq}). 
\end{itemize}

Let us see the tachyon condensation (\ref{seq}) in a little more detail.
The second arrow  in (\ref{seq}) is so-called D-brane descent relation
\cite{senconj}, and it is known that an exact tachyon profile on the
non-BPS D4-brane is given \cite{Asakawa:2002ui} by the
Atiyah-Bott-Shapiro construction \cite{Witten},
\begin{eqnarray}
t(\xi,x^i) = u \sigma_i \otimes (x^i \unit_{k\times k} - T_i(\xi)),
\label{tachyon4}
\end{eqnarray}
with the limit $u \rightarrow \infty$ ensuring it to be a classical
solution of a boundary string field theory.
The last term $T_i(\xi)$ is included to give the transverse displacement 
of the D-strings \cite{Hashimoto:2001rj}: 
the zeros of the tachyon $t$ are the location
of the topological defects (the D-strings).
To realize the first arrow in the sequence (\ref{seq}), we need the 
D-brane ascent relation \cite{Terashima:2001jc, Asakawa:2002ui} in which
we prepare infinite number of D3-branes and anti-D3-branes with the
following exact tachyon profile \cite{Terashima:2001jc}
\begin{eqnarray}
 T = u \hat{p}_\xi + it(\hat{\xi},x^i), \qquad \Phi^4 = \hat{\xi},
\label{tachyon3}
\end{eqnarray}
where again $u \rightarrow \infty$, and 
$\hat{\xi}$ and $\hat{p}_\xi$ are the infinite dimensional
matrix representation of the Heisenberg algebra 
$[\hat{\xi},\hat{p}_\xi]=i$.
The anti-hermite part of $T$ becomes the real tachyon on the non-BPS
D4-branes. 
Therefore, altogether, on the D3-branes and anti-D3-branes, if we turn
on the tachyon profile (\ref{tachyon3}) and (\ref{tachyon4}), we obtain
the desired D-string configuration with the Nahm data $T_i(\xi)$. This is
the realization of the sequence (\ref{seq}): 
\begin{eqnarray}
 T = u \hat{p}_\xi + i u\sigma_i \otimes (x^i \unit_{k\times k}
 - T_i(\hat{\xi}))
= u \nabla, \qquad \Phi=\hat{\xi}.
\label{tacdirac}
\end{eqnarray}
Quite interestingly, {\it this tachyon is identical to 
the Dirac operator}.
The viewpoint $\langle$ii$\rangle$ of the brane configuration
Fig.\ref{fig1} through the infinite number of D3-branes and
anti-D3-branes naturally gives the physical interpretation of the Dirac
operator in the Nahm's construction.\footnote{The relation
between a tachyon and the Dirac operator has
been suggested also in a probe analysis of Nahm transform 
\cite{Hori:1999me}.} 

The derivation of the Nahm's construction is completed by looking at the
interpretation $\langle$i$\rangle$, as we show in the following. 
We apply the idea of \cite{Terashima:2005ic}.
The consequence of the tachyon profile (\ref{tacdirac}) 
can be seen after it is diagonalized. The divergence of the eigenvalues 
of the tachyon matrix is directly related to the vanishing of pairs 
of the D3-brane and the anti-D3-brane. The tachyon potential of the
boundary string field theory for the
brane-antibrane system \cite{Kraus:2000nj,Takayanagi:2000rz} is given by
${\rm Tr} \exp [-u^2 {\cal D}^2]\sim {\rm Tr}\exp [-T^\dagger T]$ 
where we have defined a hermitian Dirac operator 
${\cal D}\equiv\tiny
\left(\begin{array}{cc} 0 & \nabla \\ \nabla^\dagger & 0
\end{array}\right)$.
The vanishing of the potential means the
disappearance of the D-branes \cite{senconj}.  
For the diagonalization we introduce ortho-normalized 
eigenstates of ${\cal D}$, 
${\cal D}\ket{\lambda;x^i}=e_\lambda \ket{\lambda;x^i}$.
Then the exponent of the tachyon potential is diagonally expressed as 
$u^2{\cal D}^2=\sum_\lambda
\ket{\lambda,x^i}u^2e_\lambda^2\bra{\lambda,x^i}$. 
Evidently, entries with $e_\lambda \neq 0$ diverges in the limit
$u\rightarrow \infty$. Thus only the D3-branes carrying the Chan-Paton
index of the zeromode remains, while others vanish. 

The location of the infinite number of D3-branes was originally
expressed as $\Phi = \hat{\xi}$ as in (\ref{tachyon3}).
Thus to look at the location of the remaining D3-branes, 
we just need to see particular matrix elements of $\hat{\xi}$ carrying 
the zeromode indices. They are given exactly by 
the Nahm's formula (\ref{nahmformp}), therefore the formula is derived.  

The derivation of the formula for the gauge field (\ref{nahmforma})
is a little more involved but has a fascinating interpretation:
{\it the gauge field is provided as a Berry's phase.}
In the reduction to the zeromode states, we thought of them as a
function of $x^i$, but in string theory $x^i$ 
are worldsheet boundary fields and
thus depend on the worldsheet boundary variable $\sigma$
in the path-integral formalism. 
Therefore the eigenvalue problem of the tachyon field is naturally 
accompanied by the Berry's phase, 
\begin{eqnarray}
\exp\left[
- \oint\! d\sigma \biggm\langle \!\!\rho,x^i\!\biggm| 
\frac{\p}{\p \sigma}
\biggm| \!\rho',x^i\!\!\biggm\rangle \right]
=\exp\left[
- \oint\! d\sigma \frac{\p x^i}{\p \sigma}
\biggm\langle \!\!\rho,x^j\!\biggm| \frac{\p}{\p x^i}
\biggm| \!\rho',x^j\!\!\biggm\rangle \right].
\quad
\end{eqnarray}
This phase should appear in front of the eigenstate representing the
remaining D-branes. 
Noting that $\p x^i/\p \sigma$ is nothing but the worldsheet
boundary coupling to the background gauge fields, we derive the formula
for the gauge fields, (\ref{nahmforma}). This completes the stringy
derivation of the Nahm's construction of BPS monopoles. 
We outlined the derivation here, and its details with the full rigor
is given in section \ref{sec2.1}, in the boundary state formalism
with the exact tachyon condensation. 

There exists the
``inverse'' Nahm construction in which for a given monopole field
configuration the corresponding Nahm data can be constructed. This
inverse construction can also be derived via our D-brane approach,
by employing the following new sequence:
\begin{eqnarray}
\left(
\begin{array}{c}
 \mbox{D-strings}\\
\mbox{anti-D-strings}
\end{array}
\right)
\quad\rightarrow\quad \mbox{non-BPS D4-branes}
\quad\rightarrow\quad \mbox{D3-branes}.
\label{seq2}
\end{eqnarray}
For the first arrow we need infinite number of D-strings and
anti-D-strings to realize the ascent relation, then the second arrow is
the usual co-dimension one domain wall of the tachyon field in the
descent relation. It is obvious that the same argument works,
and in fact one can derive the inverse Nahm construction. 

Originally the Nahm's construction was obtained \cite{Nahm:1979yw} 
by dimensionally reduced ADHM construction of instantons. 
In view of our derivation, it is almost obvious that the ADHM
construction itself can be derived \cite{HT} via the tachyon
condensation on D4-branes and D0-branes. D-brane realization of the
ADHM construction and Nahm transformation has been studied 
in probe D-brane analyses \cite{Douglas:1996uz,Hori:1999me} 
in background D5-D9-brane systems, but our derivation is direct
(in no need of probes) and not in the low energy approximation. 
Moreover, in our approach, the Nahm construction can be considered as
a duality: we can derive the ``inverse'' Nahm construction in the
precisely same manner as the Nahm construction, and the D-strings with
the Nahm data and the D3-branes with the monopole profile are treated 
on an equal footing. 
Note that even off-shell configurations, which do not
satisfy the BPS conditions nor the equations of motion, of the D3-branes
and that of the D-strings can be related. 

The boundary state formalism
with the exact tachyon condensation is so efficient and powerful, and we
have no doubt that it will help much for studying
generalization/application of the ADHM and Nahm constructions.  


\section{Detailed Derivation via Boundary States}
\label{sec2.1}

The equivalence of various D-brane configurations can be most
efficiently seen in terms of boundary states. Because boundary states
are considered to be
by themselves the definition of D-branes, once two boundary states
coincide, we can say that those two D-branes are identical. In the
derivation of the Nahm's construction, we make full use of the boundary
state formalism.\footnote{We employ off-shell boundary states for
exploring the tachyon condensation. Naively they have possibility of
suffering from divergences when away from on-shell background fields.
However, our main concern is the on-shell configurations 
such as monopoles,
and furthermore, the off-shell boundary states employed here have a
natural interpretation in consistency with the boundary string field
theories.}  
We will show that boundary states of the D-strings and the D3-branes
are indeed identical.

\vspace{5mm}\noindent\underline{\it Non-BPS D4 $\rightarrow$ D1}
\vspace{2mm}

First we review the second arrow in the sequence (\ref{seq}), that is,
the tachyon condensation on the non-BPS D4-branes giving a bunch of
D-strings. The final form of the sequence, 
$k$ static D-strings elongated along $\xi (\equiv x^4)$ 
direction with their transverse scalar field $T_i(\xi)$ $(i=1,2,3)$, 
can be represented by the boundary
state 
\begin{eqnarray}
\lefteqn{  |B{\rm (D1)}\rangle 
= \int [d\xi][d\psi^4] }
\nn\\
&& \hspace{5mm}
\left\{
\Tr {\rm P} 
\exp \oint d\sigma
\left[-\dot{\xi}A_\xi(\xi)
- i T^i(\xi) P_i 
+ i D_i T^j(\xi) \psi^i \Pi_j
-\frac12 [T^i(\xi), T^j(\xi)]\Pi_i \Pi_j
\right]
\right\}
\nn\\
&&\hspace{40mm}\times 
\ket{x^i=0, \xi} \ket{\psi^i=0, \psi^4}.
\label{bsd1t}
\end{eqnarray}
This is in a path-integral representation \cite{bs}, and the kets are
the eigen states of the worldsheet scalar fields, 
$X^\mu(\sigma,\tau)|_{\tau=0}\ket{x^\mu(\sigma)} 
=x^\mu(\sigma)\ket{x^\mu(\sigma)}$.
The dot denotes a derivative with respect to the worldsheet boundary 
variable $\sigma$, 
$P_\mu (\sigma)$ is the conjugate momentum of $X^\mu(\sigma)$,
and $\Pi_\mu(\sigma)$ is that of $\psi^\mu(\sigma)$. 
Tr is the trace over the Chan-Paton indices of the matrices $T_i(\xi)$,
and the path-ordering P is necessary to ensure the gauge invariance
in the target space. Hereafter we take the gauge $A_\xi=0$ for
simplicity. In this article we don't explicitly write 
the overall factor of boundary states and 
any dependence on directions different from $x^i (i=1,2,3)$ and $\xi$ in
ten dimensions, for notational simplicity. 
In this section we set $\alpha'=2$, and will argue the $\alpha'$
dependence and the low energy limits in the next section.

This boundary state (\ref{bsd1t})
can be represented in a different form, 
by using a descent relation from the non-BPS D4-branes, with the
Atiyah-Bott-Shapiro construction \cite{Witten}. 
The profile of the tachyon field $t(x^i,\xi)$ on the non-BPS D4-branes
is \cite{Hashimoto:2001rj} 
\begin{eqnarray}
t = u \sigma_i \otimes (x^i \unit_{k\times k} 
- T_i(\xi)), \qquad u \rightarrow 
+\infty.
\label{tactus}
\end{eqnarray}
This means that we need $2k$ non-BPS D4-branes to produce $k$ D-strings.
Here we set $|T_i(\xi)|=\infty$ for $|\xi|>a$, which is consistent 
because of the Nahm's boundary condition, $|T_i(\xi)|=\infty$ at
$|\xi|=a$. 
We need to take the limit $u\rightarrow \infty$ so that the resultant 
configuration is on-shell, which is shown in the boundary string field
theory. The proof of the equivalence of the boundary states 
follows precisely from the argument in \cite{Asakawa:2002ui}. 
We write here the non-BPS D4-brane boundary state as a reference,
\begin{eqnarray}
&&
 \ket{B{\rm (D4)}} =
\int [dx^i][d\xi][d\psi^i][d\psi^4]
\left\{
\Tr {\rm P}
\exp \oint d\sigma
\left[
- t^2
\unit_{2\times 2}
+ i\left(
\psi^i \p_i t + \psi^4 \p_4 t 
\right)\tau_1
\right]\right\}
\nn\\
&& \hspace{70mm} \times
\ket{x^i, \xi} \ket{\psi^i, \psi^4} ,
\label{d4m}
\end{eqnarray}
where $\tau_i$ ($i=1,2,3$) are the Pauli matrices.
One can show that (\ref{d4m}) with the tachyon profile (\ref{tactus})
is identical with (\ref{bsd1t}).\footnote{There might be a subtlety
concenring the $u\rightarrow \infty$ limit.}

\vspace{5mm}
\noindent\underline{\it D3$\overline{\it D3}$ $\rightarrow$ non-BPS D4}
\vspace{2mm}

Next, let us consider the first arrow in the sequence (\ref{seq}).
We represent the non-BPS D4-branes (\ref{d4m}) in terms of infinite
number of D3-branes and anti-D3-branes. The ascent relation 
follows from the construction in \cite{Terashima:2001jc},
according to which, the profile of the tachyon field $T(x^i)$ and the
transverse scalar field $\Phi(x^i)$ on infinite number of the 
D3-branes and the anti-D3-branes is
\begin{eqnarray}
 T = u \hat{p}_\xi + it(\hat{\xi},x^i), \qquad 
 \Phi^4 = \hat{\xi}, \qquad u\rightarrow \infty.
\label{tacd3pro}
\end{eqnarray}
Here $\hat{\xi}$ and $\hat{p}_\xi$ are matrix representation of the
Heisenberg algebra $[\hat{\xi}, \hat{p}_\xi]=i\unit$.

Although straightforward, we explicitly show here that the D3-brane (and
anti-D3-brane) boundary state with (\ref{tacd3pro}) reduces to
the non-BPS D4-brane boundary state (\ref{d4m}), for an 
instructive purpose. This is a slight 
generalization of the argument given in 
\cite{Asakawa:2002ui} where an ascent relation from a non-BPS D$p$-brane
to a BPS D$(p\!+\!1)$-brane was shown in the boundary state formalism.
We start with the D3-brane anti-D3-brane boundary state 
\cite{Kraus:2000nj,Takayanagi:2000rz} 
\begin{eqnarray}
\ket{{\rm D3}\overline{\rm D3}}
=\int [dx^i][d\psi^i]\Tr {\rm P}
\exp \left[
\oint d\sigma M(\sigma) 
\right]
\ket{\xi=0, x^i}\ket{\psi^4=0, \psi^i}
\label{bd3d3b}
\end{eqnarray}
where the matrix exponent is given by 
\begin{eqnarray}
M(\sigma) \equiv
&&
\left[
-i \Phi^4P_4(\sigma)+ i \p_i \Phi^4 \psi^i \Pi_4
\right]\unit_{2\times 2}
+
\left(
\begin{array}{cc}
- TT^\dagger & 0 \\
0 & - T^\dagger T
\end{array}
\right)
\nn\\
&&
+
\left(
\begin{array}{cc}
0 &
[T,\Phi^4]\Pi_4 + i\p_i T \psi^i \\ 
{}[T^\dagger,\Phi^4]\Pi_4 + i\p_i T^\dagger \psi^i
& 0
\end{array}
\right).
\label{msigma}
\end{eqnarray}
The matrix elements in $M(\sigma)$ can be evaluated with 
the tachyon/scalar field (\ref{tacd3pro}) as
\begin{eqnarray}
T T^\dagger =
u^2 (\hat{p}_\xi)^2 -u \p_\xi t + t^2, \quad
T^\dagger T=
u^2 (\hat{p}_\xi)^2 +u \p_\xi t + t^2, \quad 
[\Phi^4, T] =  [\Phi^4, T^\dagger] = i u,
\nn
\end{eqnarray}
which ends up with 
\begin{eqnarray}
M(\sigma) =
\left[
-i \hat{\xi}P_4 - u^2 (\hat{p}_\xi)^2 - t^2
\right]\unit_{2\times 2}
+ u \p_\xi t \tau_3
- i u \Pi_4 \tau_1
-i \p_i t  \psi^i \tau_2.
\end{eqnarray}
Then we can regard $\tau_1$ and $\tau_2$ as quantized fermions,
following \cite{Witten} ($\tau_3=i \tau_2 \tau_1$).
We introduce a boundary fermion $\psi^4(\sigma)$ replacing 
the quantized fermion $u\tau_1$ in the exponent. 
Note that the $\tau_i$ anti-commute with the fermion fields 
$\psi^i$ by definition.
With the additional fermion kinetic term which leads to the desired
quantized fermion, we obtain the expression 
\begin{eqnarray}
&& \ket{{\rm D3}\overline{\rm D3}}
 = \int [dx^i][d\psi^i]\int [d\psi^4]
\nn\\
&&
\left\{\!
\Tr {\rm P}
\exp 
\!\oint\!\! d\sigma\! 
\left[
\left(\!-\frac{\psi^4 \dot{\psi}^4}{4u^2}
\!-\!i \hat{\xi}P_4\! -\! u^2 (\hat{p}_\xi)^2 \!-\! t^2
\!-\!i \psi^4 \Pi_4\!
\right)\!\unit_{2\!\times\! 2}
+ (-\!i \p_i t \psi^i \!-\!i \p_\xi t \psi^4)\tau_2
\right]
\right\}
\nn\\
&&\hspace{70mm}\times 
\ket{\xi=0, x^i}\ket{\psi^4=0, \psi^i}.
\end{eqnarray}
Furthermore, we may replace this ``hamiltonian'' in regard of the
variables $\hat{\xi}$ and $\hat{p}_\xi$ with its lagrangian with the
path-integral over the coordinate variable $\xi(\sigma)$, following 
\cite{Asakawa:2002ui}, to get
\begin{eqnarray}
&& \ket{{\rm D3}\overline{\rm D3}}
 = \int [dx^i][d\psi^i]\int [d\psi^4]\int [d\xi]
\nn\\
&&
\left\{
\Tr {\rm P}
\exp 
\oint d\sigma 
\left[
\left(-\frac{\dot{\xi}^2\!+\!\psi^4 \dot{\psi}^4}{4u^2}
\!-\! t^2 \!-\!i \xi P_4 \!-\!i \psi^4 \Pi_4
\right)\unit_{2\times 2}
\!+\! (-\!i \p_i t \psi^i \!-\!i \p_\xi t \psi^4)\tau_2
\right]
\right\}
\nn\\
&&
\hspace{70mm}
\times 
\ket{\xi=0, x^i}\ket{\psi^4=0, \psi^i}.
\end{eqnarray}
Noting that the term $-i \xi P_4$ in the exponent just shifts 
the ket state $\ket{\xi=0}$ to $\ket{\xi}$ and so does the fermionic 
term $-i \psi^4 \Pi_4$, by taking the limit $u \rightarrow \infty$ 
we reproduce the non-BPS D4-brane boundary state (\ref{d4m}).
We had to exchange $\tau_1$ with $-\tau_2$ but this doesn't affect
anything after the trace is taken.

\vspace{5mm}
\noindent\underline{\it Nahm data $\rightarrow$ monopole}
\vspace{2mm}

Now, consider the D3-brane anti-D3-brane boundary state (\ref{bd3d3b})
and looking at the two tachyon profiles (\ref{tactus}) and
(\ref{tacd3pro}) at the same time.\footnote{We assume that $u$'s in
(\ref{tactus}) and (\ref{tacd3pro}) are the same.}
Then on the D3-branes and the anti-D3-branes,
we have a complex tachyon representing the D-strings as\footnote{
This configuration is related to the family index theorem and 
KK-theory \cite{kmatrix}.}
\begin{eqnarray}
 T = u \nabla, \qquad \nabla \equiv 
\hat{p}_\xi \unit_{2k\times 2k}
+ i \sigma_i \otimes (x^i \unit_{k\times k} 
- T_i(\hat{\xi})),
\label{tacsol3}
\end{eqnarray}
where $T_i(\xi)$ are the transverse scalars of the D-strings, 
as explained above.
This ``Dirac operator'' is exactly what appears in the Nahm's
construction.\footnote{If we have kept $A_\xi$ without gauging it away,
the derivative $\hat{p}_\xi$ in (\ref{tacsol3}) would have been replaced
by the covariant derivative $\hat{p}_\xi - i A_\xi$.} 
Substituting this with $\Phi^4 = \hat{\xi}$ into the
(\ref{msigma}), we have an expression 
\begin{eqnarray}
M(\sigma)=-i \hat{\xi}P_4 
-u^2 {\cal D}^2
+[u{\cal D}, \hat{\xi}]\Pi_4 + iu\p_i {\cal D}\psi^i, 
\label{newmsig}
\end{eqnarray}
where we have defined a hermitian operator 
\begin{eqnarray}
 {\cal D}\equiv \left(\begin{array}{cc}
0 & \nabla \\ \nabla^\dagger & 0
\end{array}\right).
\end{eqnarray}
Because this $M(\sigma)$
is in the path-ordered trace all over the representation
space, we may expand the state space in terms of the eigen states 
of the operator ${\cal D}$, 
\begin{eqnarray}
 {\cal D}\ket{\lambda, x^i} = e_\lambda \ket{\lambda, x^i}.
\end{eqnarray}
In this expression $\hat{p}_\xi$ should be understood as a derivative,
$-i\p/\p \xi$. 
A wave function can be defined as 
$\Psi_{(\lambda, x^i)}(\xi) \equiv \langle \xi\ket{\lambda, x^i}$.
Note that here $x^i$ behaves as just a parameter of the system.
The eigen states are chosen in such a way that they satisfy the
ortho-normalization condition, 
\begin{eqnarray}
\langle \lambda, x^i | \lambda', x^i\rangle
=\int\! d\xi\; \Psi_{(\lambda,x^i)}^\dagger(\xi) 
\Psi_{(\lambda',x^i)}(\xi) 
= \delta_{\lambda\lambda'}.
\end{eqnarray}
Note that $\Psi(\xi)=0$ for $|\xi| >a$ because of the 
choice $T_i(\xi)=\infty$ there.

Then, we insert a complete set 
$\sum_\lambda \ket{\lambda}\bra{\lambda}$ in the path-ordered trace
in the boundary state (\ref{bd3d3b}). Every time it hits the tachyon
part ${\cal D}^2$, there appears a factor in the trace 
\begin{eqnarray}
\exp \left[
-u^2 {\cal D}^2
\Delta \sigma
\right]
= \sum_\lambda \ket{\lambda} \bra{\lambda} 
\exp(-u^2 e_\lambda^2 \Delta \sigma).
\label{lambda2}
\end{eqnarray}
Thus in the $u \rightarrow \infty$ limit, (\ref{lambda2}) becomes
\begin{eqnarray}
\sum_\rho\ket{\rho} \bra{\rho} 
\end{eqnarray}
where $\ket{\rho}$ is a zeromode, 
\begin{eqnarray}
 {\cal D}\ket{\rho,x^i}=0.
\end{eqnarray}
Any zeromode of ${\cal D}$ is that of either $\nabla$ or
$\nabla^\dagger$. 
In the case of $k$ $SU(N)$ monopoles, 
there should appear $N$ normalizable zeromodes of $\nabla$, 
$\rho=1,2,\cdots, N$. 

Therefore, basically because of the tachyon condensation limit 
$u\rightarrow \infty$, the state sum is reduced to that for the
zeromode, and all the matrix elements in the path-ordered trace reduce
to their zeromode expectation values.\footnote{This mechanism has been
developed in \cite{Ellwood:2005yz,Terashima:2005ic}.
If the $\nabla$ part of ${\cal D}$ possesses a zeromode, correspondingly
a D3-brane remains after the tachyon condensation. On the other hand, if
a zeromode appears in the $\nabla^\dagger$ part, it means an
anti-D3-brane remaining. } 
The first term in (\ref{newmsig}) reduces to
a factor in front of the boundary state
\begin{eqnarray}
\exp \oint d\sigma 
\left[
-i \Phi(x^i(\sigma))P_4(\sigma)
\right]
\end{eqnarray}
where we define the expectation value of the operator $\hat \xi$ as 
\begin{eqnarray}
\Phi(x^i)_{\rho\rho'} \equiv \langle \rho,x^i| \hat \xi \ket{\rho',x^i}.
\label{defscalar}
\end{eqnarray}
then this $\Phi(x^i)$ forms an $N\times N$ matrix field.

In (\ref{newmsig}) the last term has only off-diagonal entries, and 
so after the overall trace is taken they appear as a pair:
\begin{eqnarray}
u^2\left(
[{\cal D}, \xi]\Pi_4 + i\p_i{\cal D}\psi^i 
\right)(\sigma_1)
\exp[- u^2 {\cal D}^2 \Delta\sigma]
\left([{\cal D}, \xi]\Pi_4 + i\p_j{\cal D} \psi^j
\right)(\sigma_2).
\label{twin}
\end{eqnarray}
Following \cite{Ellwood:2005yz}, between the two the boundary distance
$\Delta\sigma=\sigma_2-\sigma_1$ 
is fulfilled with the exponential of the diagonal elements. 
The term quadratic in $\Pi_4$ vanishes due to its fermionic feature.
Let us consider $\psi^i\psi^j$ term in (\ref{twin}) first. 
The zeromode expectation value of it is, with (\ref{lambda2}),
\begin{eqnarray}
-\sum_\lambda 
u^2\exp(-u^2 e_\lambda^2 \Delta \sigma)
\bra{\rho}
(\p_i{\cal D})
\ket{\lambda} \bra{\lambda} 
(\p_j{\cal D})\ket{\rho'}
\cdot \psi^i (\sigma_1)\psi^j(\sigma_2).
\label{elements}
\end{eqnarray}
Using relations 
\begin{eqnarray}
 \bra{\rho}(\p_i {\cal D})\ket{\lambda} = -e_\lambda
(\p_i\bra{\rho}) \ket{\lambda}, \quad 
\bra{\lambda}(\p_i {\cal D}) \ket{\rho'}
= -e_\lambda \bra{\lambda} \p_j \ket{\rho'}, 
\end{eqnarray}
and further (path-)integration over $\Delta\sigma$ 
in the $u\rightarrow \infty$ limit \cite{Ellwood:2005yz}
\begin{eqnarray}
\sum_\lambda\ket{\lambda} \bra{\lambda} 
 \int d(\Delta\sigma) u^2 e_\lambda^2 
\exp[-\Delta\sigma u^2 e_\lambda^2 ]
= \sum_{e_\lambda\neq 0}\ket{\lambda} \bra{\lambda}
= 1-\!\!\!\!\!\!\!
\sum_{\;\;\;\;\;\rho'':\;{\rm zeromode}}\!\!\!\!\!\!\!
\ket{\rho''}\bra{\rho''}, 
\nn
\end{eqnarray}
the matrix element (\ref{elements}) is evaluated as 
\begin{eqnarray}
\lefteqn{
 \left(
\left(\p_i \bra{\rho}\right) \p_j\ket{\rho'}
- \sum_{\rho''} \left(\p_i \bra{\rho}\right) \ket{\rho''}
\bra{\rho''} \p_j \ket{\rho'}
\right)\psi^i \psi^j}
\\
&&
=
 \left(
\p_i \bra{\rho}\p_j\ket{\rho'}
+ \sum_{\rho''} \bra{\rho}\p_i \ket{\rho''}
\bra{\rho''} \p_j \ket{\rho'}
\right)\psi^i \psi^j.
\end{eqnarray}
Defining the target space anti-hermitian vector field
as a zeromode expectation value of the derivative operator, 
\begin{eqnarray}
[A_i(x^j)]_{\rho\rho'}\equiv   
\langle \rho,x^j| \p_i \ket{\rho',x^j},
\label{defgauge}
\end{eqnarray}
we obtain the matrix element as 
\begin{eqnarray}
 \frac{1}{2}[F_{ij}(x^k)]_{\rho\rho'}\psi^i(\sigma)\psi^j(\sigma)
\end{eqnarray}
where the field strength is 
$F_{ij}\equiv\p_i A_j - \p_j A_i + [A_i,A_j]$.
In the same manner, terms linear in $\Pi_4$ in (\ref{twin}) gives
an exponent
\begin{eqnarray}
 iD_i \Phi (x^j) \psi^i \Pi_4.
\end{eqnarray} 

Up to now we have studied the reduction of the matrix elements
to their zeromode part, but note that the Dirac operator 
(\ref{tacsol3})
depends on $x^i$ which is actually a boundary function of $\sigma$.
Here if we regard $\sigma$ as time, then $x^i(\sigma)$ is
considered to be a time dependent parameter or an external field. 
In the limit $u\rightarrow \infty$, this dependence in the path-integral
is irrelevant (adiabatic), except for a phase factor known as Berry's
phase which is a shift for the boundary hamiltonian,
\begin{eqnarray}
 \gamma = i\int\! d\sigma 
\biggm\langle \!\rho,x^i\biggm| \frac{\p}{\p \sigma}
\biggm| \rho',x^i \!\biggm\rangle
= i\int\! d\sigma 
\langle \rho,x^i| \p_i \ket{\rho',x^i}\dot{x}^i(\sigma).
\end{eqnarray}
Thus with the definition (\ref{defgauge}), 
we obtain a bosonic worldsheet coupling to the background 
gauge field in the boundary state, 
\begin{eqnarray}
\Tr {\rm P}
\exp 
\oint d\sigma 
\left[
- \dot{x}^i A_i(x) \right],
\label{axd}
\end{eqnarray}
if we correctly incorporate the effect of the path-ordering.
A precise derivation of this term is as follows. 
First we note that
\begin{eqnarray}
\Tr {\rm P} e^{ -u^2 \oint d \sigma {\cal D}^2 (\sigma) }
\sim \Tr \prod_{l=1}^M e^{-\frac{u^2}{M} 
{\cal D}^2(\sigma_l) }, 
\label{bbb}
\end{eqnarray}
where $\sigma_l=l/M$, for $M \gg 1$.
Here we have defined the product $\prod_{l=1}^M$ with the inverse
ordering, {\it i.e.~}$\prod_{l=1}^M a_l=a_M a_{M-1} \cdots a_1$.
Then inserting 
${\rm 1}=\sum_\lambda \ket{\lambda, x^i(\sigma_l)} 
\bra{\lambda, x^i(\sigma_l)}$
and taking the $u \rightarrow \infty$ limit, (\ref{bbb}) becomes
$ \Tr \prod_{l=1}^M \sum_{\rho=1}^N
\ket{\rho, x^i(\sigma_l)} \bra{\rho, x^i(\sigma_l)}$.
Using 
$\ket{\rho, x^i(\sigma_{l-1})} \sim
\ket{\rho, x^i(\sigma_l)}- \frac{l}{M} 
\frac{\partial}{\partial \sigma_l} \ket{\rho, x^i(\sigma_l)}$,
(\ref{bbb}) is shown to be equal to (\ref{axd}).
We can easily extend this computation to our case, 
$\Tr {\rm P} e^{ -u^2 \oint d \sigma M(\sigma) }$ which is perturbed
by the terms with $P_4, \Pi_4$ or $F \psi \psi$.

Finally altogether, the full boundary state becomes the well-known
form for BPS D3-branes,
\begin{eqnarray}
&& \int [dx^i][d\psi^i]\Tr {\rm P}
\exp 
\oint d\sigma 
\left[
- \dot{x}^i A_i(x) + \frac12F_{ij} \psi_i \psi_j
-i \Phi(x^i)P_4  +iD_i \Phi (x^j) \psi^i \Pi_4
\right]
\nn \\ 
&& \hspace{8cm} \times 
\ket{\xi=0, x^i}\ket{\psi^4=0, \psi^i}.
\label{aaa}
\end{eqnarray}
Therefore, the target space fields defined in 
(\ref{defscalar}) and (\ref{defgauge}) are actually 
the scalar field and the gauge fields on the D3-branes.
We have shown that those are given by the formula 
(\ref{defscalar}) and (\ref{defgauge}), here is the completion of the
derivation of the Nahm construction of BPS monopoles.

\clearpage
\section{Some Aspects of the Derivation}

\noindent\underline{\it Low energy limits}
\vspace{2mm}

Now we study low energy limits of what we have seen.
We will find that in fact the limits are nontrivial, in the sense
that the low energy limit for the D3-branes is different from
that for the D1-branes. Nevertheless, our derivation works without 
any problem, due to the BPS nature of the configurations.

In order to see this, we should recover the $\alpha'(={l_{\rm s}}^2)$
dependence which has been taken as $\alpha'=2$ so far.
In the D-brane worldvolume actions,
it is natural to take 
the mass dimensions of the scalars $\Phi, T_i$ and 
the gauge fields $A_i$ to be unity.
We follow this and also use the convention of 
the tachyon $T$ and the parameter $u$ being dimensionless.
The separation between the most distant D3-branes is $2a$, which
means that $1/a$ has mass dimension one.
Then, the tachyon becomes
\begin{eqnarray}
 T = \frac{u{l_{\rm s}}}{\sqrt{2}} \nabla, \qquad \nabla \equiv 
\hat{p}_\xi \unit_{2k\times 2k}
+ i \sigma_i \otimes \left(\frac{2x^i}{{l_{\rm s}}^2} \unit_{k\times k} 
- T_i(\hat{\xi})\right),
\label{tacsol3a}
\end{eqnarray}
and the boundary coupling of $\Phi(x^i(\sigma))$ becomes
\begin{eqnarray}
\exp \oint d\sigma 
\left[
-i ({l_{\rm s}}^2/2) \Phi(x^i(\sigma))P_4(\sigma)
\right].
\end{eqnarray}
From these two,
we see that
\begin{eqnarray}
({l_{\rm s}}^2/2) \Phi(x^i)=\tilde \Phi (2x^i/{l_{\rm s}}^2, a),
\end{eqnarray}
where $\tilde \Phi (x^i,a)$ reduces to the previous 
$\Phi(x^i)$ when we set ${l_{\rm s}}=\sqrt{2}$,
namely $\tilde \Phi (x^i,a)$ is 
given by the original Nahm construction.
In the same manner, we can also see
\begin{eqnarray}
({l_{\rm s}}^2/2) A_i(x^i)=\tilde A_i (2x^i/{l_{\rm s}}^2,a).
\end{eqnarray}
It is important to note that
$(\Phi(x^i),  A_i(x^i))$ indeed satisfies the monopole equation
(\ref{bpseqmon}) since $(\tilde \Phi(x^i,a),  \tilde A_i(x^i,a))$
satisfies it.
Actually, by a dimensional analysis, we can show that
\begin{eqnarray}
\Phi(x^i)=\tilde \Phi (x^i, M_{D3}), \,\,\, 
 A_i(x^i)=\tilde A_i (x^i, M_{D3}),
\end{eqnarray}
where $M_{D3} \equiv 2a/{l_{\rm s}}^2 $ is the asymptotic vev of the
Higgs scalar field. 
This fact is consistent with 
our expectation that the monopole equation (\ref{bpseqmon})
is a valid BPS equation
even if we include $\alpha'$ corrections.

The Higgs vev $M_{D3}$ characterizes the 
scale of the D3-brane worldvolume theory.
On the other hand, the scale of the D1-brane worldvolume theory is set
by $M_{D1} \equiv  1/a$ since $2a$ is the length of the D1-branes
along the $\xi$ direction.
Because there is a relation $M_{D3} M_{D1} =2/{l_{\rm s}}^2$,
we can not take ${l_{\rm s}} \rightarrow 0$ limit 
keeping both $M_{D3}$ and $M_{D1}$ finite.
Therefore, if we consider the Yang-Mills theory limit 
on the D3-branes, the D1-brane worldvolume theory becomes
quite stringy with $\alpha'$ corrections non-negligible, and vice versa.

A picture based on the profile $\Phi(x^i)$, 
like Fig.~\ref{fig1}, 
is appropriate only for the low energy Yang-Mills theory.
Here the picture is supposed to represent a certain distribution,
for example the energy momentum tensor, of the D3-branes.
If we include the $\alpha'$ corrections to it, 
the picture should be modified, although
the configuration $\Phi(x^i)$ itself might not be changed by the 
$\alpha'$ corrections.
This is because the boundary state,
which includes all the $\alpha'$ corrections, with $\Phi(x^i)$
does not have the distribution localized on 
$\xi={l_{\rm s}}^2 \Phi(x^i)$, as emphasized in \cite{Terashima:2005ic}.
Note that the conditions $u \gg 1$ and $u {l_{\rm s}} /R \ll 1$ in 
\cite{Terashima:2005ic},
which are needed for the localized D-brane picture, are
$u \gg 1$ and $u {l_{\rm s}} M_{D3} \ll 1$ for our D3-branes.
This is consistent with the fact that 
the Yang-Mills limit in the D3-branes is ${l_{\rm s}} \rightarrow 0$ 
with $M_{D3}$ kept finite.
Conversely, for the D1-brane picture the condition is
$u {l_{\rm s}} M_{D1} = 2u /({l_{\rm s}} M_{D3})  \ll 1$, 
which is violated in the Yang-Mills limit 
of the D3-branes, but consistent with 
the Yang-Mills limit of the D1-branes .

The pictures based on $T_i(\xi)$ and $\Phi(x^i)$ 
look different from each other in general.
This ``difference'' is apparent especially for $k=1$ where
$T_i$ vanishes while $\Phi$ is nontrivial. 
However, these give the same boundary state and 
describe the same physical system.
This puzzle is resolved as above in view of the different low energy
limits.


\vspace{5mm}
\noindent\underline{\it Special situations in Nahm data}
\vspace{2mm}

In a general situation, the Nahm's equations (\ref{nahmeq}) include
additional terms localized on a certain point on $\xi$, so-called a
jumping data. This appears, in the D-brane interpretation,  
only when there exists a D3-brane on which the numbers of incoming and
outgoing D-strings are the same. The physical interpretation is as
follows. When those numbers are the same, there is no D-string charge 
escaping off to the spatial infinity along the $x^i$ direction, thus
it is impossible to form a D3-brane via the Myers' effect. This
``unseen'' D3-brane would be a consequence of non-normalizable
zeromodes of the Dirac operator (the tachyon profile). 
At the jumping point where the D3-brane sits, the additional term 
appearing in the Nahm equation should be provided from the hyper
multiplet coming from a string connecting the D3-brane and the
D-strings \cite{jump}. 
According to Chen and Weinberg \cite{Chen:2002vb}, this additional term
can be naturally understood to be a certain limit of the usual Nahm data
$T_i(\xi)$, when one adds another D-string so that the numbers are
different and takes the limit of bringing it off to the infinity. 
Using this ``reguralization,'' our derivation is applicable also to
the special case with the jumping points.

When we have only a single monopole $k=1$, the Nahm data vanishes: 
$T_i(\xi)=0$. It seems that the D3-brane is not existent, because with
the vanishing transverse scalar field on the single D-string it is
impossible to produce the Myers' effect. However, this apparent
contradiction can be resolved if one puts additional D-string and thinks
about a limit of bringing it off to the infinity, as above. 
One can vividly see that the Nahm data (for example provided in
\cite{Brown:1982gz}), in the limit, becomes more singular 
at the boundary points $\xi=\pm a$ where the D3-brane should appear,
while in the middle $T_i(\xi)$ approach zero.

\vspace{5mm}
\noindent\underline{\it Inverse Nahm construction}
\vspace{2mm}

One can derive the ``inverse'' Nahm construction from 
given monopole configurations $(A_i(x), \Phi(x))$ to Nahm data, 
precisely in the same manner. 
Starting with infinite number of pairs of D-strings and anti-D-strings
elongated along $\xi$, one first constructs $N$
non-BPS D4-branes by a tachyon 
profile \cite{Terashima:2001jc, Asakawa:2002ui}
\begin{eqnarray}
&& T = u (\hat{p}_i \unit_{N\times N}- i A_i(\hat{x})) \otimes \sigma^i 
+ i t(\xi,\hat{x}^i) \otimes \unit_{2\times 2}, \quad \Phi^i = \hat{x}^i,
\end{eqnarray}
and then makes curved 
D3-branes by a descent relation \cite{Hashimoto:2001rj}
\begin{eqnarray}
 t = u(\xi\unit_{N\times N}-\Phi(x^i)).
\end{eqnarray}
Altogether, we obtain a tachyon profile which is in fact the three
dimensional ``Dirac operator'' appearing in the inverse Nahm
construction,  
\begin{eqnarray}
T = u \widetilde{\nabla}, \quad \widetilde{\nabla}\equiv 
(\hat{p}_i\unit_{N\times N} - i A_i(\hat{x})) \otimes \sigma^i 
+ i (\xi\unit_{N\times N}-\Phi (\hat{x})) \otimes \unit_{2\times 2}.
\end{eqnarray}
One can show the equivalence of the D1- and D3-brane boundary states,
using this tachyon profile explicitly, and derive the formula
of the inverse Nahm construction for the Nahm data.\footnote{There
appears the Berry's phase accordingly, and this becomes a gauge field on
the D-strings, which was gauged away in the Nahm's equation
(\ref{nahmeq}).} 

\vspace{2mm}

It would be intriguing to generalize our derivation to the case of the
ADHM construction of instantons and also to the case of the
noncommutative monopoles. We would like to report on them in our
forthcoming work \cite{HT}. 

\acknowledgments 

K.~H.~would like to thank X.~g.~Chen, K.~Hori, W.~Taylor, 
P.~Yi, T.~Yoneya for comments. 
The work of S.~T.~was supported in part by DOE grant
DE-FG02-96ER40949.


\newcommand{\J}[4]{{\sl #1} {\bf #2} (#3) #4}
\newcommand{\andJ}[3]{{\bf #1} (#2) #3}
\newcommand{\AP}{Ann.\ Phys.\ (N.Y.)}
\newcommand{\MPL}{Mod.\ Phys.\ Lett.}
\newcommand{\NP}{Nucl.\ Phys.}
\newcommand{\PL}{Phys.\ Lett.}
\newcommand{\PR}{ Phys.\ Rev.}
\newcommand{\PRL}{Phys.\ Rev.\ Lett.}
\newcommand{\PTP}{Prog.\ Theor.\ Phys.}
\newcommand{\hep}[1]{{\tt hep-th/{#1}}}

\end{document}